\documentclass[%
 reprint,
nofootinbib,
 amsmath,amssymb,
 aps,
]{revtex4-1}

\usepackage{graphicx}
\usepackage{dcolumn}
\usepackage{bm}
\usepackage[normalem]{ulem}

\usepackage{epsfig,amsfonts,verbatim,enumerate}
\usepackage{txfonts}
\usepackage{graphicx,subfigure}
\usepackage[spanish,activeacute,english]{babel}
\usepackage[latin1]{inputenc}
\usepackage{latexsym}
\usepackage{framed,color}
\usepackage[colorlinks=true]{hyperref}
\definecolor{shadecolor}{rgb}{0.8,0.7,0.8}
\newcommand{\beeq}{\begin{equation}}
\newcommand{\eneq}{\end{equation}}
\newcommand{\bear}{\begin{eqnarray}}
\newcommand{\enar}{\end{eqnarray}}
\newcommand{\bef}{\begin{figure*}}
\newcommand{\enf}{\end{figure*}}
\newcommand{\kbf}{{\bf k}}
\newcommand{\xbf}{{\bf x}}
\newcommand{\qbf}{{\bf q}}
\newcommand{\rbf}{{\bf r}}
\newcommand{\pbf}{{\bf \Psi}}

\newcommand{\dd}{\mathrm{d}}

\newcommand{\zel}{Zel'dovich$\,$}
\newcommand{\mzel}{\mathrm{Zel}}
\newcommand{\mspt}{\mathrm{SPT}}
\newcommand{\msim}{\mathrm{sim}}

\def\jcap{JCAP}
\def\mnras{MNRAS}
\def\aap{A\&A}

\def\physrep{Phys.~Rep.}
\def\prd{Phys.~Rev.~D}
\def\apj{Astrophys.~J.}

\def\ssr{Space~Sci.~Rev.}

\begin{document}

\setcounter{section}{0}
\setcounter{subsection}{0}

\title{Unequal time correlators and the \zel approximation}
\author{Nora Elisa Chisari}
\affiliation{Department of Physics, University of Oxford, Denys Wilkinson Building, Keble Road, Oxford OX1 3RH, United Kingdom}
\email{elisa.chisari@physics.ox.ac.uk}
\author{Andrew Pontzen}
\affiliation{Department of Physics and Astronomy, University College London, Gower Street, London, WC1E 6BT, United Kingdom}
\email{a.pontzen@ucl.ac.uk}

\begin{abstract}
  The modeling of cosmological observables is based on the statistics of the matter density, velocity and gravitational fields in the Universe as a function of time. Typically, calculations are restricted to ``equal time'' correlations, where any given fields are evaluated at the same redshift. For some applications, it is necessary to make accurate predictions of ``unequal time correlators'', where the fields considered are evaluated at different redshifts. In this work, we show that the \zel approximation provides an accurate ($<10\%$) analytical prescription to model unequal time correlators, which we validate against numerical $N$-body simulations. The \zel approximation introduces a scale-dependent exponential suppression of unequal time correlators, which depends on cosmology and the redshifts of the fields considered. Comparing the \zel case to previous approximations, we show that it can yield accurate predictions for wavenumbers that extend well into the nonlinear regime. However, we also show that correlations over such scales are typically suppressed by the geometry of the lightcone, and thus should normally be negligible for cosmology with galaxy surveys. We discuss potential exceptions, such as intrinsic galaxy alignments, where unequal time correlators could play a role in the modeling of the observables. 
\end{abstract}
\maketitle

\section{Introduction}
\label{sec:intro}

The extraction of information from cosmological large-scale structure relies on accurate modeling of the distribution of matter. As the next decade brings telescopes such as the Large Synoptic Survey Telescope\footnote{\url{http://www.lsst.org}} \citep{LSST}, {\it Euclid}\footnote{\url{sci.esa.int/euclid/}} \citep{Laureijs11} and {\it WFIRST-AFTA}\footnote{\url{http://WFIRST.gsfc.nasa.gov}} \citep{green11} to first light, the statistical uncertainties in measurements of large-scale structure observables, such as the clustering of galaxies or weak gravitational lensing, will decrease drastically compared to existing surveys. This will be partly driven by the increased depth of these surveys, partly by extended area coverage. In this context, it becomes necessary to refine our models for these observables in order to meet survey accuracy goals.

A key input in the modeling of the large-scale structure is the power spectrum of the observed fields, i.e. the correlation of their Fourier space components. This statistic is most often computed for fields that are evaluated at equal time, or equal redshift. However, realistic measurements involve integration along the line-of-sight, and so require prediction of correlations between cosmological fields at different redshifts, or ``unequal time correlators''. These predictions are typically made through different approximations.

The ``geometric approximation'' is a common approach to modeling full-sky observables of galaxy lensing and  clustering \citep{Kilbinger17,Angpow}; intrinsic galaxy alignments \citep{Kirk12,Chisari13}; and, in general, any two-point correlation functions of density field tracers \citep{Schmidt12}. Consider the Fourier transform of the matter over-density field, $\delta(\kbf,z)$, at wavenumber $\kbf$ and redshift $z$. We would like to compute the cross-spectrum between the density field at two different redshifts, $z$ and $z'$. In the geometric approximation, this is simply given by a geometric mean of the auto-spectra at those redshifts, i.e.,
\begin{equation}
  P^G(\kbf,z,z')\simeq\sqrt{P(\kbf,z)P(\kbf,z')}.
  \label{eq:geo}
\end{equation}
Moreover, it is often assumed that the evolution of $P$ with redshift can be obtained by re-scaling the matter power spectrum today by the growth function of matter perturbations, $D(z)$, for a given cosmology and redshift, reducing Eq. (\ref{eq:geo}) to
\begin{equation}
  P^{GL}(\kbf,z,z')\simeq D(z)D(z')P_L(k).
  \label{eq:geolin}
\end{equation}
where $P_L(k)$ is the linear matter power spectrum at $z=0$.

Few works have investigated the accuracy of this approximation to model unequal time correlators. The authors of \citep{Kitching17} have proposed to use Eulerian (standard) perturbation theory (SPT) \citep{Goroff86,Makino92,Jain94,Bernardeau02} to obtain a more accurate prediction of $P(\kbf,z,z')$, compared to the linear case. However, SPT is known to be accurate only below $k\simeq 0.2-0.3\,h\,{\rm Mpc}^{-1}$ for equal time correlators \citep{McQuinn16,Vlah16}.

The \zel approximation \citep{Zeldovich70} is the linear order Lagrangian perturbation theory (LPT), and it is known to be exact up to shell crossing in one dimension. Numerous works have described the application of the \zel approximation to two-point redshift-space clustering, most of them focusing on reconstruction and modeling baryon acoustic oscillations \citep{Eisenstein07a,Eisenstein07b,Castorina18}. Extensions to $N$-point functions have also been performed \citep{Tassev14a}. LPT (either the \zel approximation or the second-order 2LPT) is also used to set up initial conditions of such simulations \citep{Knebe09,Eisenstein16,Garrison16} and even to correct for general relativistic effects in $N$-body Newtonian simulations \citep{Chisari11}. For further applications, see \cite{White14} and references therein. 

In this work, we show that the \zel approximation can provide an accurate approximation to unequal time correlators across a wide range of scales and redshifts, improving substantially over SPT. The reason for this is the ability of the \zel approximation to capture advection by large-scale flows in the Universe \cite{Tassev12,Tassev14b}. In Reference~\cite{Schmittfull18} it was recently suggested the use of LPT for correcting for the effect of bulk flows on the comparison between simulations and perturbative predictions from initial conditions at the field level (``shifted operators''). We demonstrate the accuracy of such an approximation in this work. For unequal time correlators, we find that the relative displacement of matter between redshifts introduces an exponential suppression of the small scale modes, which allows us to extend the regime of validity of the predictions into the fully nonlinear regime. Nevertheless, we will also show that the characteristic wavelength at which the exponential suppression becomes significant is orders of magnitude smaller than the equivalent suppression wavelength from projection along the line of sight. This establishes that, for many observable quantities, the existing geometric linear ansatz~\eqref{eq:geolin} is likely sufficient.

In Section \ref{sec:zel}, we introduce our formalism and develop an expression for $P(\kbf,z,z')$ in the \zel approximation. We compare this approximation to the modeling of the same quantity in SPT in Section \ref{sec:spt}. We present our main results in Section \ref{sec:res}, where we compare our the different predictions to measured matter power spectra in cosmological $N$-body simulations in Section \ref{sec:comp_sims} and we analyse the differences between SPT and the \zel approximation in Section \ref{sec:comp_spt}. We discuss specific applications of unequal time correlators for cosmology from large-scale surveys in Section \ref{sec:apply}, where we show that their contribution tends to be highly suppressed in the lightcone geometry. Our conclusions are presented in Section \ref{sec:conclusion}. 

\section{\zel approximation}
\label{sec:zel}

In the \zel approximation, the density field of the Universe at a given time can be obtained by displacing the initial Lagrangian positions of particles, $\qbf$, by means of a displacement function, $\pbf(\qbf, z)$, such that the final positions are: $\xbf(\qbf, z)=\qbf+\pbf(\qbf, z)$. As a consequence of this transformation, the Fourier components of the over-density field take the form
\begin{equation}
  \delta(\kbf,z) = \int \dd^3\qbf\, e^{-i\kbf\cdot\qbf}\left[e^{-i\kbf\cdot\pbf(\qbf)}-1\right]\textrm{,}
\end{equation}
where the displacement function $\pbf$ transforms the field from the initial conditions to redshift $z$.

In \citep{Pontzen16}, the \zel approximation was used to derive the cross-spectrum between two numerical simulations with inverted initial conditions. Here, we closely follow their approach to derive an expression for unequal time correlators of the matter power spectrum. Consider the density field of the universe at two different redshifts, $z$ and $z'$. The cross-spectrum between these two fields, $P(\kbf,z,z')$, is defined as
\begin{equation}
  \langle \delta(\kbf,z)\delta(\kbf',z') \rangle = (2\pi)^3 \delta^{D}(\kbf+\kbf')P(k,z,z')\textrm{,}
\end{equation}
where angle brackets indicate an ensemble average and $k=|\kbf|$. The Dirac delta, $\delta^{D}$, is a consequence of translational invariance. In the \zel approximation, this cross-spectrum depends on the relative displacement of the particles between the two redshifts: $\Delta\pbf(\qbf,\qbf'; z,z') = \pbf(\qbf,z)-\pbf(\qbf',z')$. Given that $\langle\delta(\kbf,z)\rangle=0$, we can write the cross-power explicitly as
\begin{equation}
P(k,z,z')  =
\int \dd^3\qbf\, \dd^3\qbf'\, e^{-i\kbf\cdot(\qbf-\qbf')}
\left[\left\langle e^{-i\kbf\cdot\Delta\pbf}\right\rangle-1\right].
\label{eq:cross1}
\end{equation}

Assuming that the displacements are small, they can be modeled from linear theory as proportional to the linear density field at $z=0$, $\delta_L(\kbf,z=0)$, via
\begin{eqnarray}
\pbf  = i \frac{\kbf}{k^2} \delta_L(\kbf,z=0) D(z). 
\label{eq:zapp}
\end{eqnarray}
An analogous relation holds for $\pbf'$. In this linear-displacement limit, $\pbf$ is Gaussian distributed. The ensemble average $\langle e^{-G}\rangle$ of a Gaussian field $G$ satisfies the cumulant relation $\langle e^{-G} \rangle = e^{-1/2\langle G^2\rangle}$, which we can substitute in Eq. (\ref{eq:cross1}). Following a similar derivation to that of Appendix A of Ref. \citep{Pontzen16}, we arrive at the following expression for the unequal time correlator in the \zel approximation,
\begin{align}
 P^{\mzel}(\kbf,z,z') &= \int \dd^3\rbf\, e^{-i\kbf\cdot \rbf}\times \nonumber \\
&\hspace{1.0cm} \left[e^{-\left[D(z')-D(z)\right]^2I(\kbf,0)/2+D(z')D(z)J(\kbf,\rbf)}-1\right]
\label{eq:a8}
\end{align}
where we have defined two auxiliary functions:
\begin{eqnarray}
  I(\kbf,\rbf)&=& \int\frac{\dd^3\kbf'}{(2\pi)^3}\frac{(\kbf\cdot\kbf')^2}{k'^4}\cos(\kbf'\cdot\rbf)P_L(k'),\\
  \textrm{ and } J(\kbf,\rbf)&=& I(\kbf,\rbf)-I(\kbf,0) .
\end{eqnarray}
Notice that the constant term in the integrand of Eq.~\eqref{eq:a8} integrates to a $\kbf=0$ correction that we will ignore. 
We can rewrite the $\rbf$-dependent part of Eq.~\eqref{eq:a8} as the Zel'dovich predicted power spectrum
$P^{\mzel}(\kbf,\bar{z})$ at an intermediate redshift $\bar{z}$:
\begin{equation}
  \int d^3\rbf\, e^{-i \kbf \cdot \rbf - D(z')D(z)J(\kbf,\rbf)}
  =  P^{\mzel}(\kbf,\bar{z},\bar{z}) \equiv  P^{\mzel}(\kbf,\bar{z})\label{eq:zeldovich-NL-autopower}
\end{equation}
where $\bar{z}$ is defined to satisfy $D(z')D(z) = D^2(\bar{z})$, guaranteeing that $\bar{z}$ is intermediate between $z$ and $z'$. Finally the
 term proportional to $I(\kbf,0)$ in the exponential within Eq. (\ref{eq:a8}) is independent of $\rbf$ and can be pulled out of the integral. As a consequence, there is an overall exponential suppression factor in the expression for $P^{\mzel}(k,z,z')$ which is $k$-dependent:
\begin{equation}
  P^{\mzel}(k,z,z')=P^{\mzel}(k,\bar{z}) \, e^{-\left[D(z')-D(z)\right]^2(k/k_{\rm NL})^2} \textrm{,}
  \label{eq:a10}
\end{equation}
and valid for $k \ne 0$. Here we have defined a nonlinear scale 
\begin{equation}
  k_{\rm NL}^{-2} = \frac{1}{12\pi^2}\int_0^\infty P_L(k') \dd k'.
  \label{eq:kNL}
\end{equation}
Note that this scale is calculated using $P_L$, the linear matter power spectrum, at $z=0$, since the linear growth is normalized to $D(0)=1$.

If we re-linearize \eqref{eq:zeldovich-NL-autopower} by Taylor expanding the exponential to first order in the quantity $D(z')D(z)$ we recover the geometric linear ansatz, Eq.~(\ref{eq:geolin}); but the full result for the cross-spectrum, Eq.~\eqref{eq:a8}, retains the exponential suppression for sufficiently high $k$. Thus, at sufficiently high redshift, we recover 
\begin{align}
  P^{\mzel}(k,z,z') & \to  P^{GL}(k,z,z')\,e^{-\left[D(z')-D(z)\right]^2(k/k_{\rm NL})^2} \nonumber \\
 \textrm{ as } D(z) & \to  0 \textrm{ and } D(z') \to 0 \textrm{.}
  \label{eq:putccomp}
\end{align}

This result illustrates why the $P^{GL}(k,z,z')$ approximation has proved successful in part to model unequal time correlators. At sufficiently large $z$ and for sufficiently low $k$, the approximation is recovered exactly. On the other hand, the \zel approximation introduces a scale- and time-dependent correction to the geometric linear approximation which is present at sufficiently high $k$ even when $D(z)$ is small. The suppression in cross-power is a consequence of the relative displacement of the matter field between two redshifts due to gravity; the typical wavenumber at which it becomes important is $k_{\mathrm{NL}}/\left|D(z')-D(z)\right|$. The suppression factor reduces to unity when $z= z'$, as expected. 

It is valid to question whether either of the expressions~\eqref{eq:a10}~or~\eqref{eq:putccomp} can be useful given that the corrections become important in a high-$k$, non-linear regime where the auto-power will not be well approximated either by $P_L(k)$ or $P^{\mzel}(k)$. Empirically, we will show in Section~\ref{sec:res} that our resummed \zel approximation gives an excellent fit to the ratio between cross- and auto-power spectra from fully non-linear simulations, all the way to small scales $k \sim 10\,h\,\mathrm{Mpc}^{-1}$. This is despite the fact that the prediction for the auto-power spectrum delivered by \zel is not accurate enough for cosmological applications \citep{Vlah15}. The reason why \zel works so well for our purposes must therefore be attributed to taking ratios such that the overall growth of non-linear structure is factored out. Once this ratio is taken, high-$k$ power suppression is driven by displacements arising from near-linear low-$k$ velocity fields. As emphasized by Ref.~\cite{Pontzen16}, accurate power spectrum ratios can be recovered in this limit through a careful choice of resummation. 

\section{Standard perturbation theory}
\label{sec:spt}

Previous work by Kitching \& Heavens \citep{Kitching17} proposed the use of standard perturbation theory for modeling unequal time correlators. In this context, the nonlinear matter overdensity field is expressed as a perturbative expansion over a set of functions $f_n(\kbf)$, which are functions involving $n$ powers of $\delta_L(\kbf,z=0)$. The perturbative expansion is expressed as follows,
\begin{equation}
  \delta(\kbf,z) = \sum_{n=1}^\infty D^n(z)f_n(\kbf)
  \label{eq:dspt}
\end{equation}
where $D(z)$ is the growth factor we have defined in Section \ref{sec:intro}. The expansion of the matter power spectrum up to quartic order in the density (the ``one-loop'' approximation) at a given redshift is 
\begin{equation}
  P^{\mspt}(k,z)=D^2(z)P_L(k)+D^4(z)P_{22}(k)+2D^4(z)P_{13}(k)
  \label{eq:pkptone}
\end{equation}
where $P_{22}$ and $P_{13}$ involve the auto-correlation of $f_2$ and cross-correlation of $f_1$ and $f_3$ terms respectively \citep[e.g.][]{McQuinn16}.

Reference \citep{Kitching17} extended the application of the expansion in Eq. (\ref{eq:dspt}) to find an approximate expression for the unequal time power spectrum at one-loop order in SPT \citep[Eq. (6) in][]{Kitching17}, 
\begin{align}
  P^{\mspt}(k,z,z') &= D(z)D(z')P_L(k)+D^2(z)D^2(z')P_{22}(k) \nonumber \\
& \hspace{0.5cm}+[D^3(z)D(z')+D(z)D^3(z')]P_{13}(k).
  \label{eq:pkuetcpt}
\end{align}
The first term in Eq. (\ref{eq:pkuetcpt}) recovers the geometric linear approximation, Eq. (\ref{eq:geolin}). We will also refer to this term as ``11'' to follow standard notation in SPT. The SPT expression incorporates two new terms which are scale- and time-dependent. The second term of the sum, we will refer to as ``22''; and the third, as ``13''. 

A shortcoming of SPT is that the sum of $P_{22}(k)$ and $P_{13}(k)$ terms tends to cancel out at large wavenumbers, leading to unstable numerical predictions. Regularizations of these terms \citep{FASTPT} are known to stabilize the calculation in the presence of these large cancellations. In the case of unequal time correlations, the re-scaling of each term by powers of the growth factor break the exact cancellation, allowing for better numerical convergence (Eq. \ref{eq:pkuetcpt}). Nevertheless, we expect SPT to be most successful below a typical $k \lesssim 0.2-0.3\,h/$Mpc. We will see in Section \ref{sec:comp_spt} that the \zel approximation  provides a  more stable and accurate expression for unequal-time power spectrum calculation.

\section{Results}
\label{sec:res}

\subsection{Comparison to simulations}
\label{sec:comp_sims}

\begin{figure*}[!ht]
  \centering
  \includegraphics[width=0.75\textwidth]{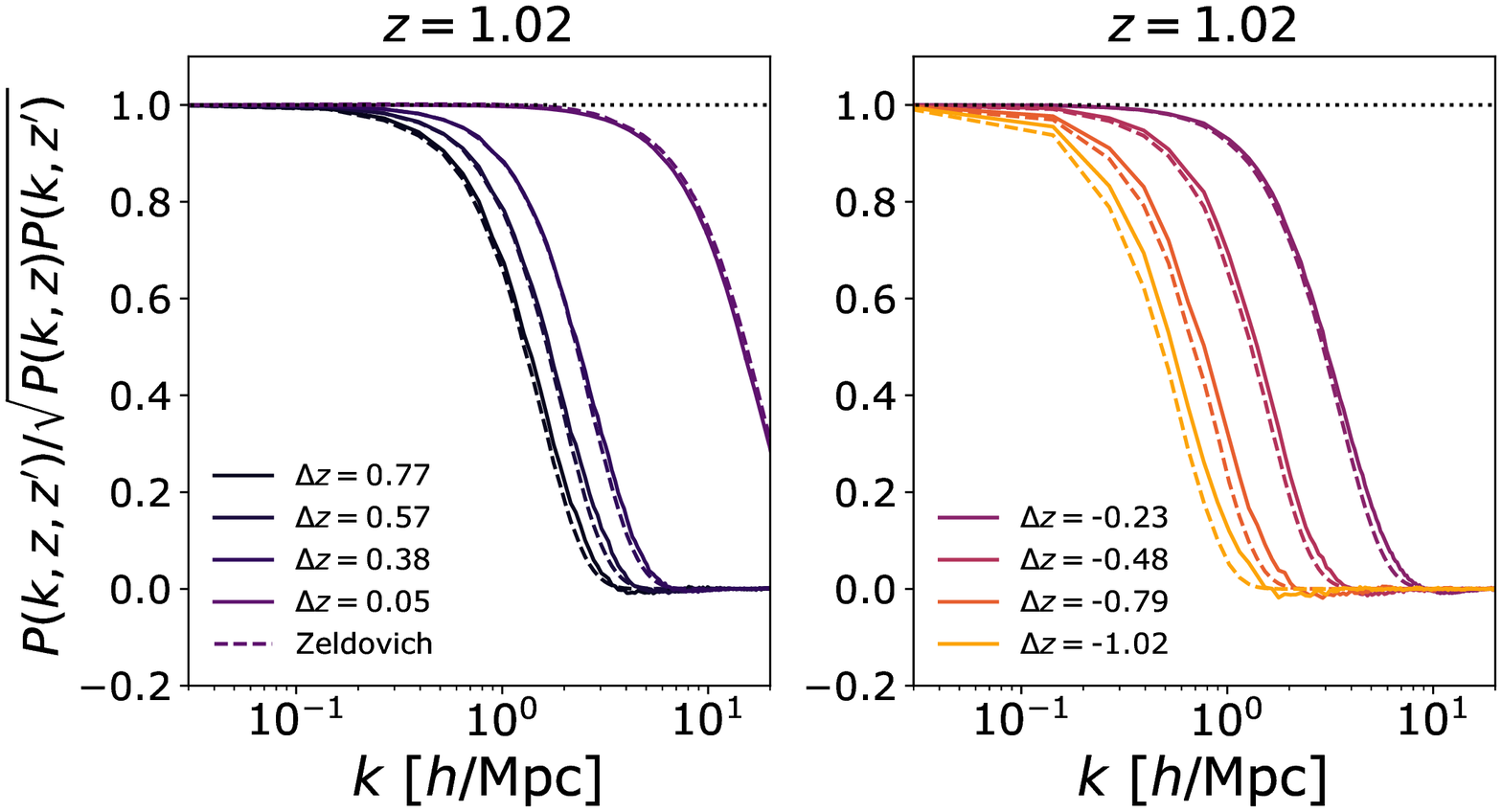}
  \includegraphics[width=0.75\textwidth]{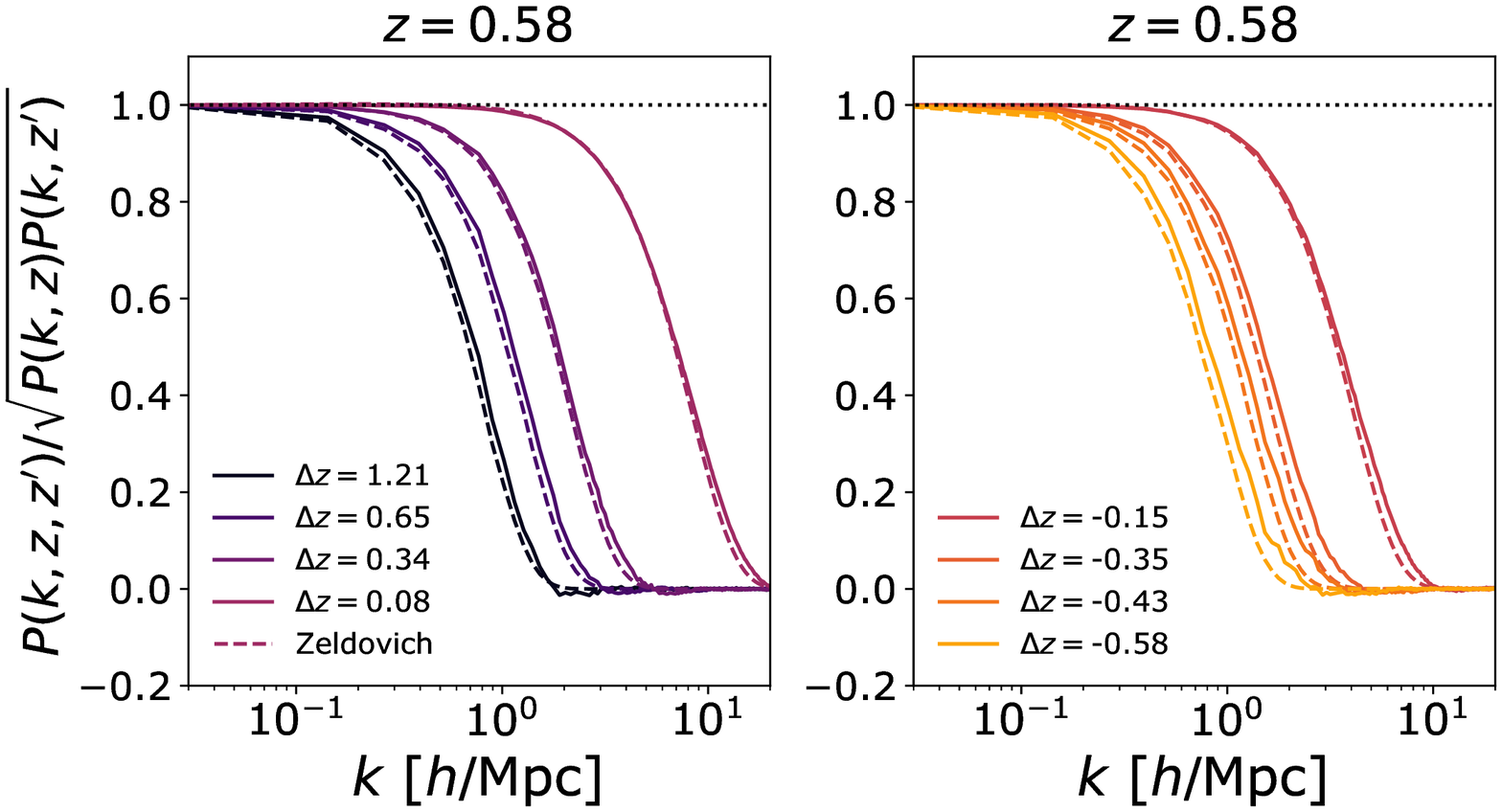}
  \caption{The predicted ratio between the cross-spectrum of the density field at $z$ and $z' = z + \Delta z$ and the geometric mean of the auto-spectra. Specifically, we compare $r_{\msim}$ (Eq. \ref{eq:rsim}, solid lines) and $r_{\rm Zel}$ (Eq. \ref{eq:rzel}, dashed lines). The top panels adopt $z=1.02$ as a reference redshift; while the bottom panels use $z=0.58$. These are intended to represent the typical mean redshift of future and current weak lensing surveys, respectively. From left to right, we vary $\Delta z=z'-z$ in the range $0\leq z'<2$. Overall the \zel approximation, which to leading order is a simple exponential suppression of small-scale power given by Eq.~\ref{eq:rzel2}, accurately describes the decorrelation of density fields due to dynamical evolution. The horizontal dotted line in each panel represents the geometric mean (linear theory) prediction.}
  \label{fig:allsnaps}
\end{figure*}

To assess the accuracy of the \zel approximation for modeling unequal time correlators, we perform a set of comparisons between the predictions derived in the previous sections and numerical simulations. To this end, we use a cosmological $N$-body simulation of $200\,h^{-1}$ Mpc on each side ran with the smoothed-particle-hydrodynamics code GADGET \citep{GADGET2,GADGET1}. The cosmology adopted for this run is consistent with current constraints from the cosmic microwave background obtained by the {\it Planck} collaboration \citep{Planck16}. The configuration of cosmological parameters used corresponds to a flat $\Lambda$CDM universe with a dark matter density of $\Omega_c=0.26$, a baryon density of $\Omega_b=0.05$, a primordial spectral index of $n_s=0.96$, a Hubble constant of $H_0=67.27\,{\rm km}/{\rm s}$ ${\rm Mpc}^{-1}$ and $\sigma_8=0.831$. The dark matter particle mass was $M_{\rm DM}=7.8\times 10^9 {\rm M}_\odot$. The simulation outputs span the range $0\leq z \leq 2$, which are typically of interest to future lensing surveys. All equal and unequal time power spectra of the simulated matter density field were obtained by using the publicly available software {\tt genPk} \citep{genPk}.

We tested the accuracy of the \zel approximation by comparing the simulated cross-spectra, $P^{\msim}(k,z,z')$, to our analytical prediction derived in Section \ref{sec:zel}. We normalize $P^{\msim}(k,z,z')$ by the more commonly adopted geometrical approximation, effectively showing our results in terms of the ratio
  \begin{equation}
    r^{\msim} \equiv \frac{P^{\msim}(k,z,z')}{\sqrt{P^{\msim}(k,z)P^{\msim}(k,z')}}.
    \label{eq:rsim}
  \end{equation}
  where all power spectra are computed directly from the simulations.
  To assess the accuracy of the \zel approximation, we compare $r^{\msim}$ to
  \begin{equation}
    r^{\rm Zel} \equiv \frac{P^{\msim}(k,\bar{z})e^{-\left[D(z')-D(z)\right]^2(k/k_{\rm NL})^2}}{\sqrt{P^{\msim}(k,z)P^{\msim}(k,z')}},
    \label{eq:rzel}
  \end{equation}
i.e.  the unequal time power spectrum estimate is obtained by using Eq. (\ref{eq:a10}), replacing the \zel auto-power spectrum with $P^{\msim}(k,\bar{z})$. As a reminder, $\bar{z}$ is obtained by numerically solving the requirement that $D^2(\bar{z})=D(z)D(z')$. One could alternatively adopt
  \begin{equation}
   r'_{\rm Zel} \equiv e^{-\left[D(z')-D(z)\right]^2(k/k_{\rm NL})^2}.\label{eq:rzel2}
  \end{equation}
  In practice, the difference between $r_{\rm Zel}$ and $r'_{\rm Zel}$ is $\lesssim 5\%$ for all $z$ and $z'$ considered and in the $k<1.5\,h$ Mpc$^{-1}$ range, where $r$ is significantly above zero. For the adopted cosmology, we also estimated $k_{\rm NL}=0.24\,h$ Mpc$^{-1}$ from Eq. (\ref{eq:kNL}).

\begin{figure*}[ht]
  \centering
  \includegraphics[width=0.49\textwidth]{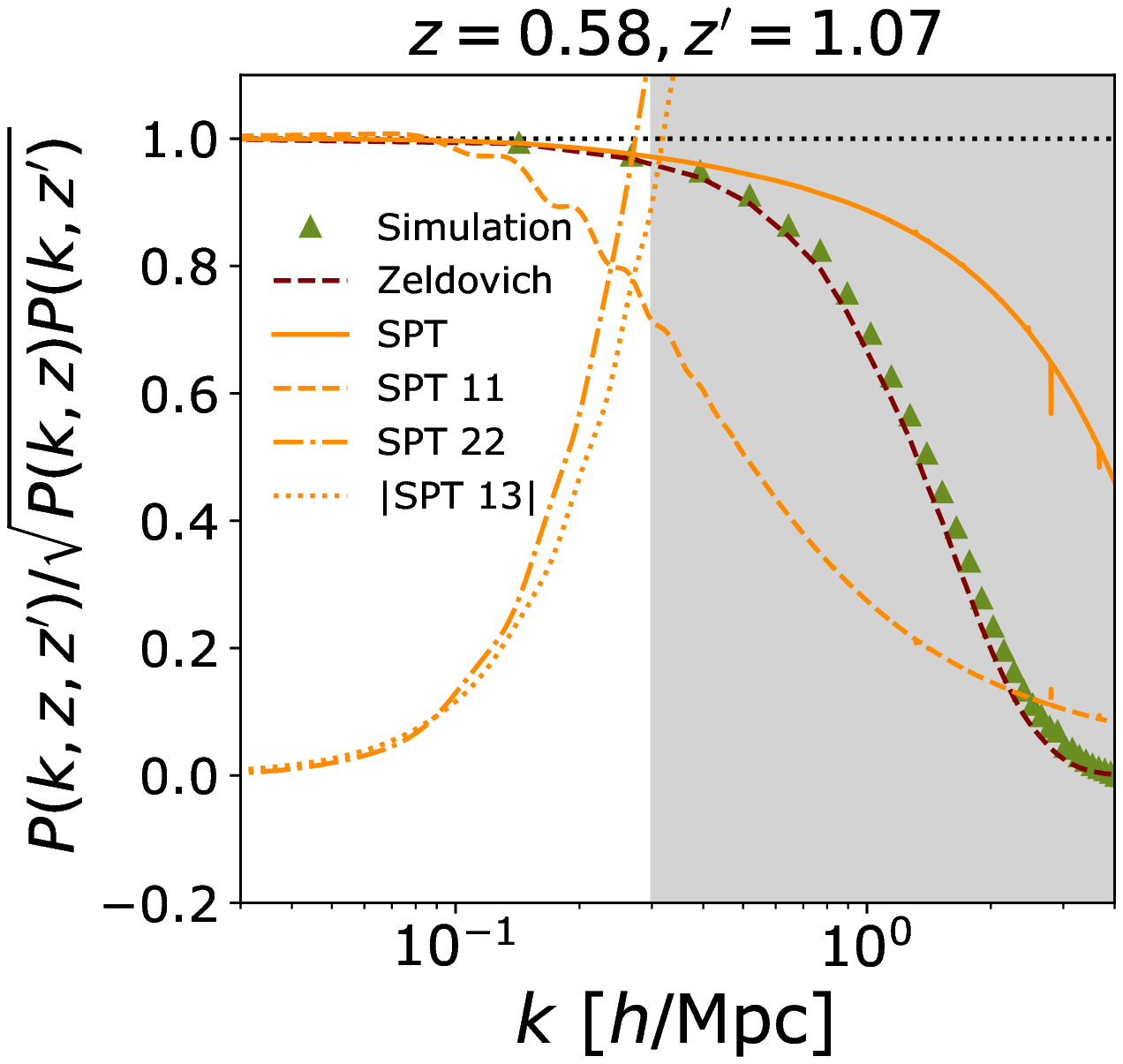}
  \includegraphics[width=0.49\textwidth]{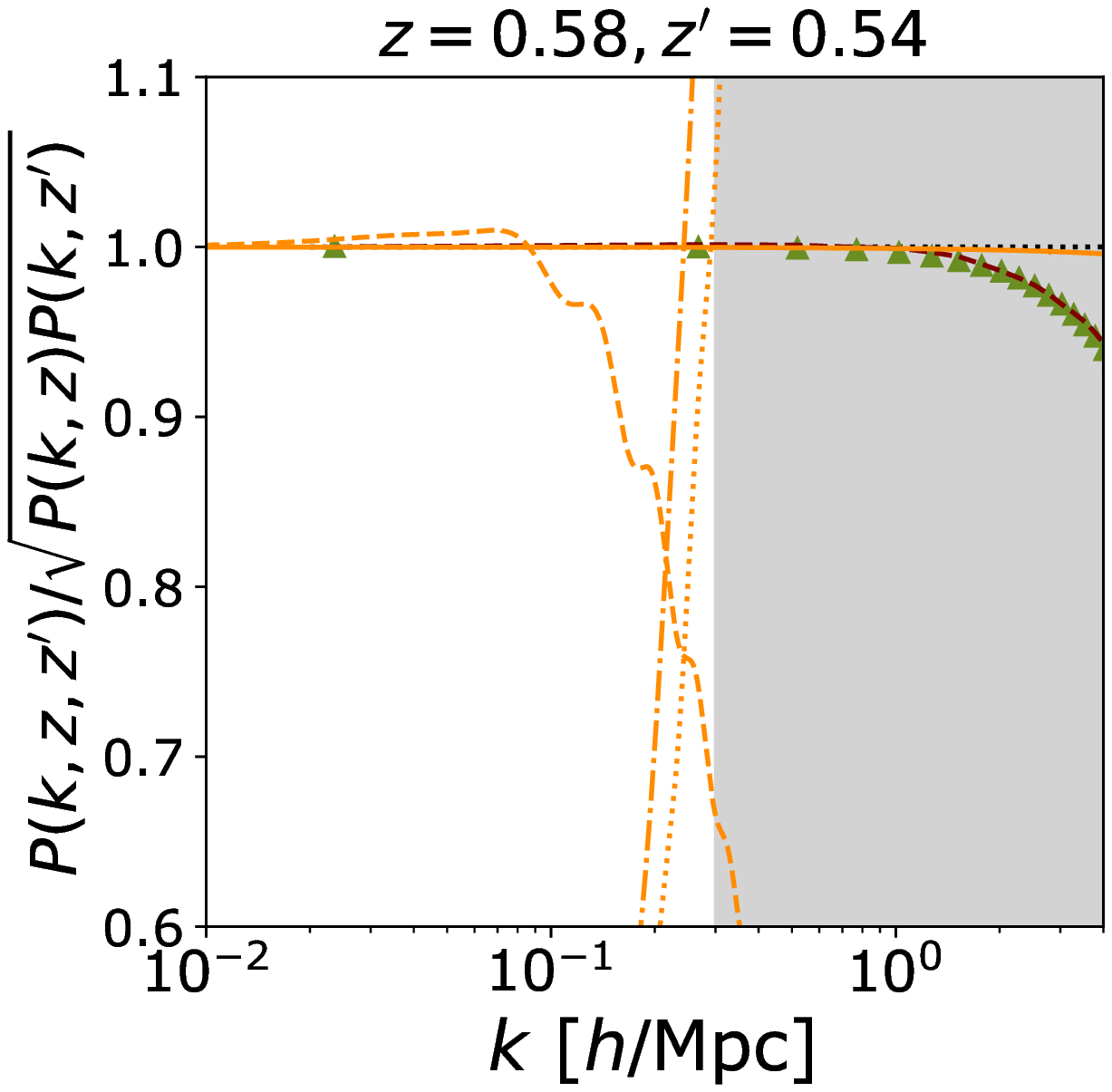}
  \caption{The predicted ratio between the cross-spectrum of the density field at $z$ and $z'$ and the geometric mean of the auto-spectra, including the SPT case. The left panel corresponds to $z'=1.07$ and $z=0.58$, and the right panel, to $z'=0.54$ and $z=0.58$. In this figure, the simulation is represented by the green triangles ($r_{\msim}$, Eq. \ref{eq:rsim}), the \zel approximation by the red dashed line ($r_{\rm Zel}$, Eq. \ref{eq:rzel}) and the SPT predictions are shown in orange. The solid orange line represents the sum of the different SPT terms that contribute up to second order on the density field. Each of the terms is shown in orange with different line styles: $11$ (dashed), $22$ (dot-dashed), the absolute value of $13$ (dotted). SPT predictions are accurate up to $k=0.3\,h/$Mpc depending on the specific application. Note that for the \zel prediction, we are plotting Eq. (\ref{eq:a10}), while for the simulations we divide the simulated cross-spectrum between the two redshifts by the geometric mean of the simulated auto-spectra. For SPT, we normalize by the geometric mean of the SPT prediction.}
  \label{fig:simcomp}  
\end{figure*}

We chose two reference redshifts $z=\{1.02,0.58\}$ for which to obtain $P^\msim(k,z,z')$, which represent the typical mean redshift of the next and current generation of weak lensing surveys, respectively. We then sampled $z'$ over the range of interest for these surveys, considering both large and small redshift separations. These two limits are representative of the typical redshift uncertainty due to photometric redshifts (i.e. $\Delta z \equiv |z-z'| \simeq 0.04$) and of the separation between typical tomographic bins (i.e. up to $\Delta z \simeq 1$). For each $z'$, we show in Figure~\ref{fig:allsnaps} the comparison between $r_{\msim}$ (solid lines) and $r_{\rm Zel}$ (dashed lines). The results demonstrate that the \zel approximation gives an excellent prediction for the ratio of the cross-spectrum to the geometric mean for both reference redshifts and throughout the whole range of $z'$ considered. The exponential factor introduced by the \zel approximation is crucial to account for large-scale flows, as discussed at the end of Section~\ref{sec:zel}. The geometric linear mean approximation (black dotted horizontal lines) systematically and significantly overpredicts the cross-power spectrum at $k\gtrsim 1\,h$ Mpc$^{-1}$. The error in the suppression factor, given by $r_{\msim}-r_{\rm Zel}$, is limited to $<10\%$ (and is typically of order $5\%$) for all $z$ and $z'$, even up to $k=10\,h\,{\rm Mpc}^{-1}$. This is likely to be accurate enough for any practical purposes given that the suppression factor anyway constitutes only a small correction to observables (see Section \ref{sec:apply}).

Notice that we have not, at any point, attempted to validate predictions of the overall amplitude and scale dependence of the matter power spectrum in SPT nor in the \zel approximation. The modeling approach presented in this work aims for successfully describing $r_{\msim}$ alone.

\subsection{Comparison to standard perturbation theory}
\label{sec:comp_spt}

We compute the SPT prediction for the ratio of the unequal time power spectrum to the geometric mean by defining $r_{\rm SPT}$ in complete analogy with Eq. (\ref{eq:rsim}). We compare this prediction to $r_{\msim}$ and $r_{\rm Zel}$ in Figure \ref{fig:simcomp}. To obtain the SPT one-loop power spectra, we use the publicly available {\tt pkd}\footnote{\url{http://wwwmpa.mpa-garching.mpg.de/\textasciitilde komatsu/}} software \citep{Jeong06} for computing $P_{22}(k)$ and $P_{13}(k)$, similarly to Ref. \citep{Kitching17}.

We choose $z=0.58$ and consider the large-difference case ($z'=1.07$, left panel), as well as a more typical photometric redshift bin width ($z'=0.54$, right panel). We see that the \zel approximation is much more accurate than SPT in both limits for predicting the ratio between the cross-spectrum of the density field and the geometric mean of the auto-spectra at $z$ and $z'$. 

\section{Applications}
\label{sec:apply}

The \zel approximation to unequal time power spectra can be applied to the modeling cosmological observables when the redshift range is wide, or when two fields that are being cross-correlated are separated by a large redshift baseline. For example, the \zel approximation could provide an alternative avenue for modeling the unequal time power spectrum for weak lensing \citep{Kilbinger17} and clustering \citep{Angpow} in photometric surveys. However, the contribution of the unequal time correlator competes with the pure geometric effect of projection along the lightcone. In other words, in a lightcone geometry, widely separated redshift bins are not expected to show correlations except on very long scales, where the linear approximation the power spectrum may be sufficiently accurate. Since we have shown dynamical effects introduce exponential suppression of the cross-power beyond a characteristic wavenumber, we now turn to estimating the equivalent wavenumber for projection effects.

\subsection{Competition with projection effects}

Let us temporarily ignore all dynamical effects discussed above and consider the impact of projection effects alone. We define the three-dimensional correlation function of the density field at a given redshift and as a function of the projected comoving radius, $r_{\perp}$, and line-of-sight comoving distance, $\Pi$, as
\begin{equation}
  \xi_{\delta \delta}(r_\perp,\Pi,z) = \int \frac{dk_\perp^2 dk_z}{(2\pi)^3}\,P(k,z)\,e^{i(k_\perp\cdot r_\perp+k_z\Pi)},
  \label{eq:xidd}
\end{equation}
where $k^2=k_\perp^2+k_z^2$. The projected 2D correlation function $w(r_\perp, z)$ is then given by the integral over the line of sight: $w(r_\perp,z)=\int d\Pi\,  \xi_{\delta \delta} $. Within a finite line-of-sight interval $[-\Pi_{\rm max},\Pi_{\rm max}]$, this can be expanded to give
\begin{equation}
w(r_\perp,z) = \frac{1}{\pi^2}\int_0^\infty dk_\perp\,k_\perp\,J_0(k_\perp r_\perp)\int_0^\infty dk_z\,\frac{\sin(k_z\Pi_{\rm max})}{k_z}P(k,z).
\end{equation}
We thus define a projected\footnote{Note that an alternative and closely-related quantity of interest is the cross-power between two thin slices separated by $\Pi_{\rm max}$. We verified that characteristic decorrelation scale and conclusions below are unaltered by considering this alternative quantity, which is obtained by removing the $k_z$ denominator and replacing $\sin$ by $\cos$ in the integrand of equation~\eqref{eq:pk2d}.} two-dimensional power spectrum analogue as
\begin{equation}
  P_{\rm 2D}(k_\perp,\Pi_{\rm max},z)=\int_0^\infty dk_z\,\frac{\sin(k_z\Pi_{\rm max})}{k_z}P(k,z).
  \label{eq:pk2d} 
\end{equation}

\begin{figure}
  \includegraphics[width=0.49\textwidth]{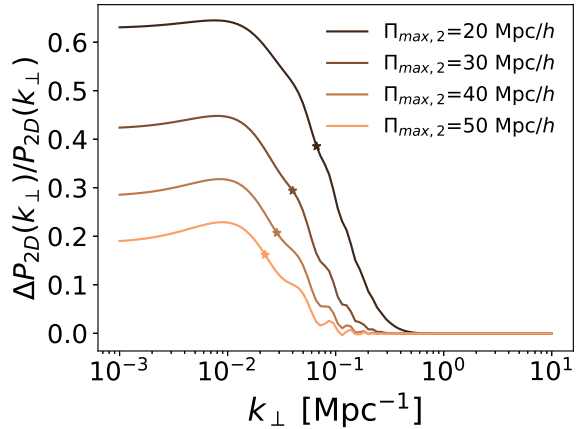}
  \caption{Geometric decorrelation of cosmological 2D power spectra for successive comoving line-of-sight separations in the absence of the dynamical effects. Each line shows the ratio $\Delta P_{\rm 2D}(k_\perp)/P_{\rm 2D}(k_\perp)$ for successive $10$ $h^{-1}$ Mpc slices in line-of-sight distance. Farther away slices contribute less to the projected  two-dimensional power spectrum and, at high $k_\perp \gg 1/\Pi_{\rm max}$, stop contributing any power at all. The stars indicate the point at which $k_\perp =1/\Pi_{\rm max}$, justifying the use of this relation as a characteristic decorrelation scale associated with projection or geometry.}\label{fig:geom-decorrelation}
\end{figure}

To study the impact of the projection, let us start by defining a hypothetical survey with a $20\,h^{-1}\,{\rm Mpc}$ line-of-sight bin, i.e. with a projected power spectrum ${P_{\rm 2D}(k_\perp) \equiv P_{\rm 2D}(k_\perp,\Pi_{\rm max}=10\,h^{-1}\,{\rm Mpc})}$. 
We will centre our survey on $z=0.58$, and suppress this $z$ dependence from the notation for brevity. Next, we consider increasing the width of the survey bin incrementally in $10$ $h^{-1}$ Mpc slices, and ask how much additional power is introduced to the projected density distribution by these increasingly distant regions. We define the increment of the two-dimensional power spectrum at each step as
\begin{equation}
\Delta P_{\rm 2D}(k_\perp, \Pi_{\rm max})=P_{\rm 2D}(k_\perp,\Pi_{\rm max})-P_{\rm 2D}(k_\perp,\Pi_{\rm max} - 10 \, {\rm Mpc}/h).
\end{equation}
Figure \ref{fig:geom-decorrelation} shows the ratio $\Delta P_{\rm 2D}(k_\perp, \Pi_{\rm max}) / P_{\rm 2D}(k_\perp)$ for $\Pi_{\rm max} = 20, 30, 40$ and $50\,{\rm Mpc}/h$. In other words, each line shows the additional 2D power introduced as the notional survey bin is broadened along the line-of-sight. The most important effect is that contribution to the observable power is {\it exponentially suppressed} beyond $k_{\perp} = 1/\Pi_{\rm max}$. (In addition, the overall amplitude of $\Delta  P_{\rm 2D}(k_\perp)$ also decreases as the slices become separated more widely.)

To understand why projected power is rapidly suppressed in the limit where $k_\perp \Pi_{\rm max} \gg 1$, note that the integrand of the projected power~\eqref{eq:pk2d} is highly oscillatory. The point $k_\perp = 1/\Pi_{\rm max}$ is indicated in Figure~\ref{fig:geom-decorrelation} by stars, and indeed we can see that the exponential decorrelation takes hold at this typical scale.

\begin{figure}
  \includegraphics[width=0.49\textwidth]{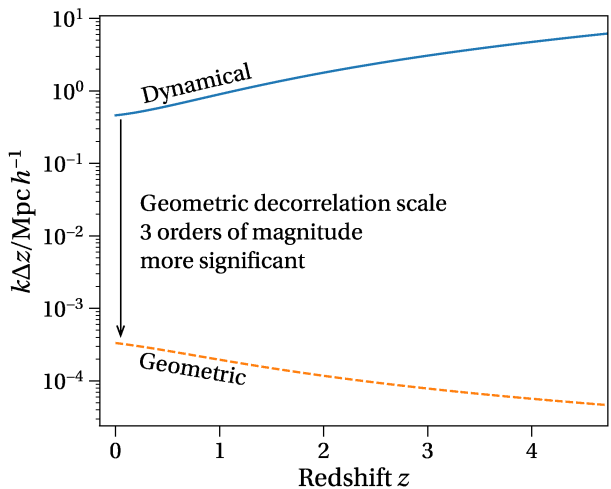}
  \caption{Comparison of the pure geometric and dynamical decorrelation scales, where the latter is computed using the \zel approximation and the former is given by a rough estimate described in the text. The pure geometric decorrelation always applies at vastly larger scales and therefore dominate over dynamical effects for most conceivable observables. }\label{fig:scale-comparison}
\end{figure}

To compare this projection decorrelation scale with the dynamical decorrelation scale calculated via the \zel approximation, we require to convert both into redshift space. For the projection effects, if the two observed line-of-sight slices are spaced by a sufficiently small redshift interval $\Delta z$, the corresponding comoving distance is $c\Delta z/H(z)$, and thus the decorrelation wavenumber associated with pure geometry is
\begin{equation}
k_{\mathrm{geom}} \Delta z = \frac{H(z)}{c}\textrm{.}
\end{equation}
On the other hand, from dynamical structure growth as approximated by equation~\eqref{eq:a10}, the decorrelation wavenumber is
\begin{equation}
k_{\mathrm{dyn}} \Delta z = k_{\mathrm{NL}} \left(\frac{\dd D(z)}{\dd z}\right)^{-1}\textrm{.}
\end{equation}
These two estimates are compared in Figure \ref{fig:scale-comparison} to show that the dynamical effects always apply at vastly larger wavenumbers (i.e. much smaller scales) than pure geometrical decorrelation. Thus we should expect geometric and projection effects to be more important, by orders of magnitude, than dynamical effects.

The suppression factor is roughly equivalent to the ratio between the free-fall velocities from structure formation to the speed of light. This is because pure projection effects decorrelate cross-spectra on scales comparable to the line-of-sight separation of two slices. In agreement with this heuristic picture, Figure~\ref{fig:scale-comparison} shows that, at high redshift, the separation of scales becomes progressively stronger and dynamical effects therefore matter even less. Of course this is a vastly simplified model of true observations, so our result does not rule out that dynamical unequal-time effects have some role to play in precision cosmology ---  but it does suggest that they should always be subdominant. 

The commonly adopted Limber approximation \citep{Limber53} essentially makes use of the geometric decorrelation when assuming that modes contributing to cosmological observables are typically transverse to the line of sight \citep{Blandford91}. In the phrasing of the unequal time correlator, the contribution of $P(k,z,z')$ to observables can be neglected when $z\neq z'$. However, this assumption fails at large scales ($k \ll k_{\mathrm{geom}}$), as correlation is preserved, and thus full-sky observables need to be modeled beyond the Limber approximation. In that regime, we have seen that the geometric linear model for the matter power spectrum, Eq.~(\ref{eq:geolin}), becomes accurate.

\subsection{Are there any circumstances in which unequal time correlators matter?}

The intrinsic correlations of galaxy shapes \cite{Troxel15,Joachimi15} are a known contaminant to weak lensing cosmology. Reference \cite{Hirata04} demonstrated that it becomes crucial to model the cross-correlation between the matter field responsible for gravitational lensing, and the intrinsic shapes of galaxies sourced by the tidal field of the large-scale structure. Intrinsic alignments have been observed to high significance in current surveys, most recently in the LOWZ sample of the Baryon Acoustic Oscillation Survey \citep{Singh15} and they have been proposed as a probe of cosmology in their own right \citep{Chisari13,Chisari14,Schmidt15,Chisari16}.

One of the main uncertainties in the modeling of intrinsic alignments is their time evolution \cite{Kirk12}. Current models vary in their assumptions of when galaxies acquire their alignment properties. One case assumes that galaxies react instantaneously to the effect of large-scale tides (although this hypothesis is under pressure from theoretical arguments \cite{Camelio15}). Other scenarios assume that galaxies establish their alignment at some earlier redshift and evolve passively thereafter (see discussion in \cite{Blazek15}). In this second type of scenario, the \zel approximation can be a viable alternative to model the correlation between the matter field at a given redshift and the alignment sourced by the tidal field at a given earlier epoch. We plan to explore this avenue of study in future work.

In recent work, reference \cite{Schmitz18} made predictions of intrinsic alignment observables assuming that galaxies established their alignment at an initial position and redshift and evolved passively thereafter. In their approach, the \zel approximation was applied in the computation of the advection contribution, while Eulerian SPT was used to model the cosmological fields. Our work therefore gives some indirect support to the approach of  Ref. \cite{Schmitz18}: we have established that a low-order Lagrangian description correctly captures the dynamical decorrelation of power even on highly non-linear small scales.

\section{Conclusions}
\label{sec:conclusion}

In this work, we have presented an application of the \zel approximation to model unequal time correlators for cosmology which we have validated by performing a comparison to numerical $N$-body simulations. We presented a simple recipe to obtain accurate predictions for unequal time correlators by re-scaling predictions for the nonlinear matter auto-spectra, which can be obtained from fits to simulations or emulators, for example.

We have also discussed the application of the \zel approximation to unequal time correlators in the context of future photometric galaxy surveys, and we have shown that the impact of unequal time correlators should be subdominant due to projection effects in the lightcone geometry. Nevertheless, unequal time correlators might be of use in the modelling of certain cosmological observables such as galaxy intrinsic alignments; in this context our work has established that low-order Lagrangian ``advection" correctly describes the decorrelation of power even at high wavenumbers, in the deeply non-linear regime.

\vspace{1cm}

\acknowledgments

We thank Anze Slosar, Tom Kitching, Alan Heavens, Marcel Schmittfull and Jonathan Blazek for useful discussions. NEC is supported by a Royal Astronomical Society Research Fellowship. AP is supported by the Royal Society, and by STFC Consolidated Grant number ST/R000476/1. This work was partially enabled by funding from the University College London (UCL) Cosmoparticle Initiative.

This work has made use of the Core Cosmology Library {\tt v1.0.0} software \citep{CCL}, a publicly released LSST DESC product available via \url{https://github.com/LSSTDESC/CCL}.

\bibliographystyle{apsrev4-1}
\bibliography{paper}

\end{document}